\documentclass{emulateapj}

\shorttitle{Bar Effects on Central Star Formation and AGN Activity}
\shortauthors{Seulhee Oh et al.}

\begin{document}
\title{Bar Effects on Central Star Formation and AGN Activity}
\author{Seulhee Oh, Kyuseok Oh and Sukyoung K. Yi}
\affil{Department of Astronomy and Yonsei University Observatory, Yonsei University, Seoul 120-749, Republic of Korea; yi@yonsei.ac.kr}

\begin{abstract}
Galactic bars are often suspected to be a channel of gas inflow to the galactic center and trigger central star formation and active galactic nuclei (AGN) activity. However, the current status on this issue based on empirical studies is unsettling, especially on AGN. We investigate this question based on the Sloan Digital Sky Survey (SDSS) Data Release 7. From the nearby ($0.01 < z < 0.05$) bright ($M_{\rm r} < -19$) database, we have constructed a sample of 6,658 relatively face-on late-type galaxies through visual inspection. We found 36\% of them to have a bar. Bars are found to be more common in galaxies with earlier morphology. This makes sample selection critical. Parameter-based selections would miss a large fraction of barred galaxies of early morphology. Bar effects on star formation or AGN are difficult to understand properly because multiple factors (bar frequency, stellar mass, black-hole mass, gas contents, etc.) seem to contribute to them in intricate manners. In the hope of breaking these degeneracies, we inspect bar effects for fixed galaxy properties. Bar effects on central star formation seem higher in redder galaxies. Bar effects on AGN on the other hand are higher in bluer and less massive galaxies. These effects seem more pronounced with increasing bar length. We discuss possible implications in terms of gas contents, bar strength, bar evolution, fueling time-scale, and the dynamical role of supermassive black hole. 
\end{abstract}

\section{Introduction} 

Galactic bars are considered to a product of secular evolution and found in all morphological types of disk galaxies. About 60$\%$ of bright disk galaxies show bar structures in the near-infrared (Knapen 2000; Eskridge 2000). 

For normal disk galaxies spiral arms contain more dust than bulges do, and naturally star formation is more pronounced in disks than in bulges. However, sometimes intense star formation are found in circumnuclear region, and active galactic nuclei (AGN) activities are also frequently observed.
Gas is the most important element in both star formation and AGN activity. The issue is how gas has been supplied to the relatively gas-poor central regions.  Numerical simulations based on gas dynamics give a clue to bar-induced fueling in disk galaxies (Sanders \& Huntley 1976; Combes \& Gerin 1985; Shlosman, Frank, \& Begelman 1989; Athanassoula 1992; Friedli, Benz, \& Kennicutt 1994; Englmaier \& Gerhard 1997; Ann \& Thakur 2005). According to these simulations, stellar orbits and gas flow patterns pass near the galactic center along an elongated orbit within a bar. So, bars efficiently drive outer gas to the circumnuclear region of galaxies.

Evidence for this theoretical expectation has been sought for by numerous studies. First, regarding the bar effect on central star formation, barred galaxies show enhanced radio and far-infrared emissions compared with unbarred galaxies (e.g. Hummel 1981; Hawarden et al. 1986; Devereux 1987). The star formation rate (SFR) is also found to be higher in barred galaxies than in unbarred galaxies (Hummel et al. 1990; Martin 1995; Huang et al. 1996; Ellison et al. 2011). Viewing from a different aspect, several studies have found an excess of bar galaxies among the galaxies with central star formation (e.g., Heckman 1980; Arsenault 1989; Huang 1996; Ho, Filippenko, \& Sargent 1997; Hunt \& Malkan 1999; Hao et al. 2009). 

The relation between large-scale bar and AGN activity is still unclear. Simkin, Su, \& Schwarz (1980) found a higher frequency of AGN among barred galaxies, while Fricke \& Kollatschny (1989) could not find such an excess of Seyfert nuclei among barred galaxies. Studies comparing bar galaxy fractions between active (in terms of black-hole activities) and inactive galaxies show contradictory results. Some studies found a higher frequency of bars among AGN (Arsenault 1989; Moles, M$\acute{\rm a}$rquez, \& P$\acute{\rm e}$rez 1995; Knapen, Sholsman, \& Peletier 2000; Laurikainen, Salo, \& Buta 2004; Hao et al. 2009), while others could not confirm it (Mulchaey \& Regan 1997; Hunt \& Malkan 1999; Martini et al. 2003).

Numerical simulations suggest that the efficiency of driving gas along a bar is sensitive to {\em bar strength} (e.g. Athanassoula 1992; Friedli \& Benz 1993; Regan \& Teuben 2004). Observational evidence does not appear to be conclusive so far. Using far infrared luminosities as a proxy for SFR, Pompea \& Rieke (1990) and Isobe \& Feigelson (1992) found no evidence of bar strength effect, whereas Aguerri (1999) found a positive correlation. Ellison et al. (2011) argued that the relation found by Aguerri (1999) may have originated from the mass-luminosity relation. Moreover, {\em statistical analyses} for the effect of bar strength also produced conflicting results. Some found an excess of star-forming galaxies or AGN in strong barred galaxies (Martin 1995; Ho et al. 1997), but others did not (M$\acute{\rm a}$rquez et al. 2000).

Most statistical tests on the bar effects are performed by comparing bar galaxy fractions between active and inactive galaxies. However, several studies have found that the bar fraction is non-monotonic and varies with galaxy morphology (Odewahn 1996; Elmegreen, Elmegreen, \& Hirst 2004; Giordano et al. 2010; Nair \& Abraham 2010; Masters et al. 2011). So, comparing bar fractions without considering galaxy properties can be highly affected by the sampling bias. Moreover, both central star formation and AGN activity have a strong connection with galaxy properties such as morphology, color, black-hole mass, and so on. Galaxy colors correlate with galaxy morphology and are determined by its star formation history. The stellar mass can be an indicator of galaxy size and gravitational potential. The supermassive black hole is probably a main driver of AGN activity.

Most of the results in this paper will be shown by comparing statistics between barred and unbarred galaxies. We then compare the fractions and emission line strengths of central star formation and AGN activity between barred and unbarred galaxies for fixed galaxy properties. This will pin down the net bar effect on central star formation and AGN activity, thereby, minimizing sampling bias. Bar length have a positive correlation with bar strength (Block et al. 2004; Elmegreen et al. 2007), so in this study we examine the effect of bar length for central star formation and AGN.

Previous studies were limited to a small number of galaxies in the nearby universe. A much larger sample is now available from large-scale surveys. We use the Sloan Digital Sky Survey (SDSS) database. Throughout this study, we assume a $\Lambda$CDM cosmology with $\Omega_{0} = 0.27$, $\Omega_{\Lambda} = 0.73$, $\sigma_{8} = 0.77$, $\Omega_{\rm b}/ \Omega_{0}=0.17$, and H$_{0} = 71$ km s$^{-1}$ Mpc$^{-1}$.

\section{Data construction}

\subsection{The Sloan Digital Sky Survey}

\begin{figure*}[t]
\centering
\includegraphics[width=0.9\textwidth]{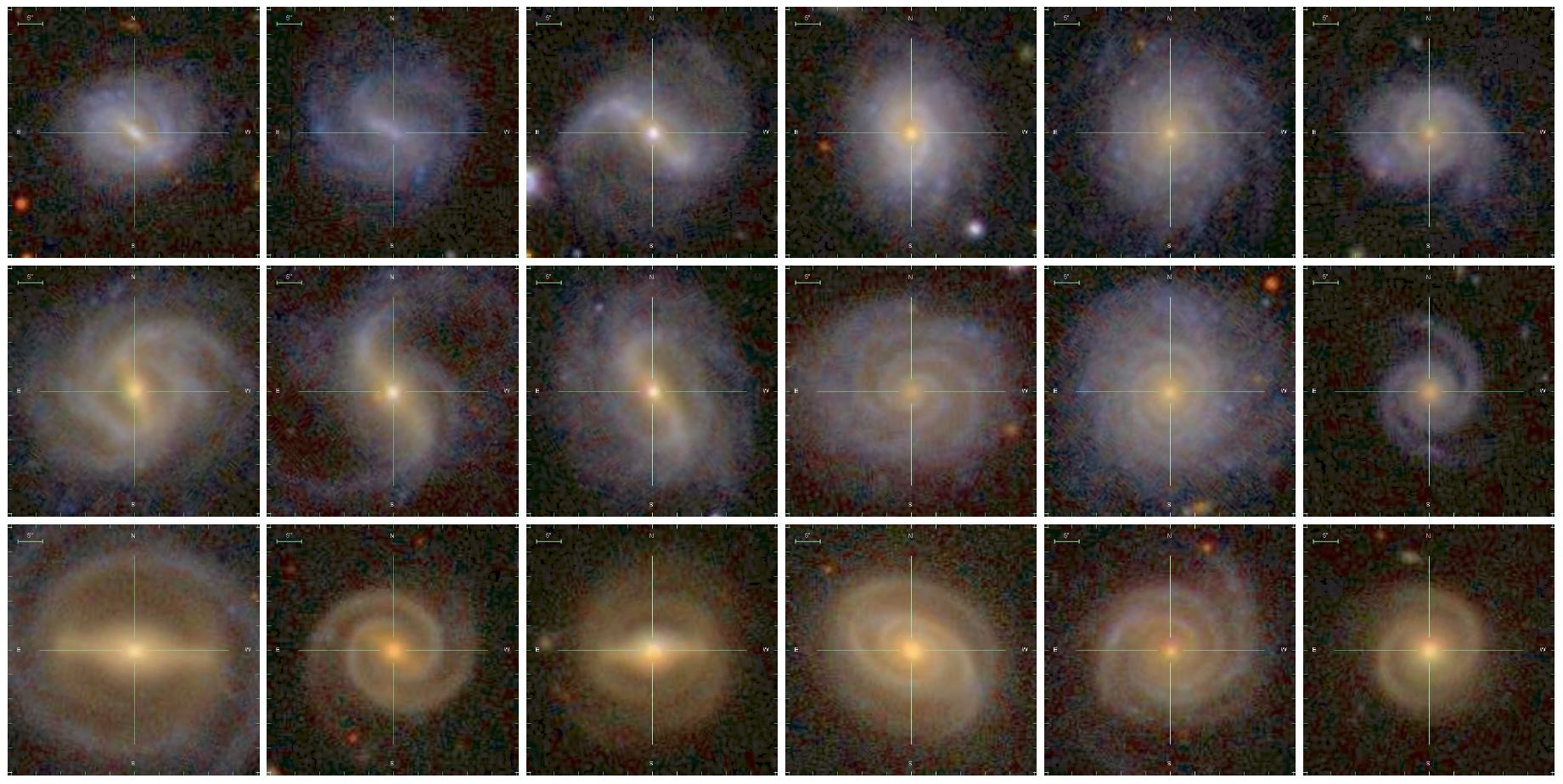}
\caption{Sample SDSS color-composite images of our late-type galaxies. Three columns on the left and right are examples of barred and unbarred galaxies, respectively. Each image covers 50$''$ $\times$ 50$''$ and a 5$''$ ruler is shown in each panel.}
\label{sample}
\end{figure*}

We used the ``main'' galaxy sample of the SDSS Data Release 7 (SDSS DR7; Abazajian et al. 2009), which is the final data release of SDSS-II project. The SDSS spectroscopic survey employs optical fibers with a 3$''$ diameter (York et al. 2000; Blanton et al. 2003). For our sample galaxies, the SDSS fiber typically covers the central 10\% in diameter, reasonably tracing much more confined ``central'' activities, albeit at a diluted level. 

We get photometric and spectroscopic data from the Catalog Archive Server (CAS) for the DR7. 
In this study, we use petrosian magnitudes (Petrosian 1976) for estimating galaxy luminosity. 
On the other hand, we use model magnitudes for colors because they give more reliable measures of unbiased colors for extended sources (Strauss et al. 2002). 

We correct foreground Galactic extinction by using dust maps provided by the SDSS pipeline (Schegel, Finkbeiner \& Davis 1998). We conduct $k$-correction using the algorithm $\bold{kcorrect\;v4_{-}1}$ (Blanton \& Roweis 2007) on magnitudes and colors.

\subsection{Sample selection}

\begin{table}[t]
 \begin{center}
  \caption[Summary of sample selection]
  {Summary of sample selection}
  \begin{tabular}{ll}
  \hline \hline
 \multicolumn{1}{l}{Criterion} & Explanation \\
 \hline
  $0.01 < z < 0.05$ & Redshift range for reliable morphological \\  & classification without saturation\\
 $M_{\rm r} < -19.$ & The absolute $\rm r$-band magnitude cut for\\  & volume limited sample\\
 $IsoB_{\rm r}/IsoA_{\rm r}$ $\ge$ 0.7 & Exclude edge-on galaxies \\
 Visual inspection & A selection of late-type galaxies\\ & which enable to classify their morphology\\
   \hline \hline \\
\end{tabular}
\label{tab:sample}
\end{center}
\end{table}

We use galaxies in the redshift range $0.01 < z < 0.05$. The lower cut is to avoid saturated images and the upper cut is to ensure the accuracy of morphology inspection. The SDSS spectroscopic limiting magnitude, $r =17.77$ corresponds to $M_{\rm r} = -19.0$ at redshift 0.05. Edge-on galaxies are not ideal for our study because both morphological classification and bar detection/measurement are difficult. Besides, it is difficult to interpret spectroscopic data for highly inclined galaxies because line strengths are line-of-sight integrated. So, we use only the galaxies with an apparent axial ratio $IsoB_{\rm r}/IsoA_{\rm r} \ge 0.7$, where $IsoA_{\rm r}$ and $IsoB_{\rm r}$ are isophotal semi-major and semi-minor axes of galaxies in the $r$-band. This axial ratio limit corresponds to $45^\circ$ in terms of inclination.

We then perform visual inspection on all candidate galaxies using color-composite images in order to construct a sample of late-type galaxies. We chose visual classification over index-based ones (e.g., colors or concentration index) to minimize selection bias. The significance of our choice and sampling strategy will be discussed in Section 5.

For late-type galaxies, we checked the presence of a bar from optical images and classified them into ``barred'' or ``unbarred'' galaxies. We kept only the galaxies for which we could perform reasonably safe morphology classification. There are 94,519 galaxies in the redshift range $0.01 < z <0.05$ in the SDSS DR7 from which 28,008 galaxies remain after initial cut. Among them, we chose 6,658 robust late-type galaxies with decent image qualities. We detected a bar in 2,422 galaxies (36\%) of them. Figure~\ref{sample} shows sample galaxies with and without a bar.

\begin{figure}[t]
\centering
\includegraphics[width=0.45\textwidth]{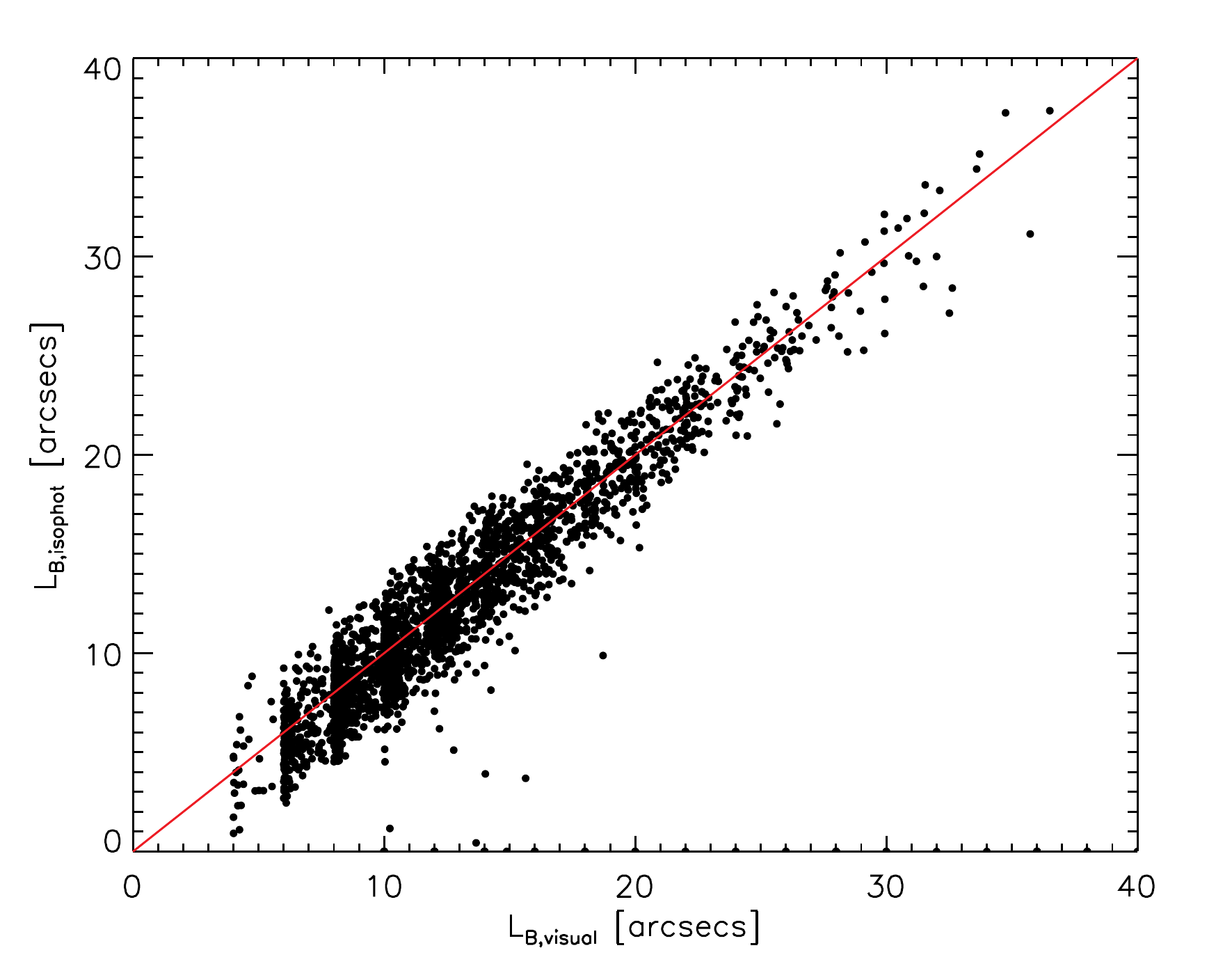}
\caption{Comparison of bar lengths derived from visual measurement and ellipse fitting (in $''$). Both of them are corrected for the projection effect. The red line indicates one-to-one correlation. Vertical features are due to the unit length of visual measurement, which is $2''$. Bar lengths from the two different methods are in good agreement with each other.}
\label{leng}
\end{figure}

\section{Analysis}
\subsection{Measurement of bar length}
For barred galaxies, we directly measured bar lengths from the SDSS color-composite images in units of 2 $arcsecs$. We felt during the measurement exercise that we could not measure bar lengths any better than this. Although we eliminated high-inclination galaxies through the initial sampling cut, we cannot totally escape from the effects of inclination on the bar length measurement. We use the following formula of Martin (1995) to correct for the projection effect:
\begin{eqnarray}
L_{\rm Bar} = l_{\rm Bar}\,(cos^2 \theta + sec^2i \,sin^2\theta)^{1/2},
\end{eqnarray}
where $l_{\rm Bar}$ is the measured length of the bar, $\theta$ is the position angle between the semi-major axis of the bar and the disk, and $i$ is the inclination of the galaxy. 

The position angle of a bar is determined by using the ELLIPSE task within the STSDAS package in IRAF (Image Reduction and Analysis Facility; http:// iraf.noao.edu/). Ellipse fitting is performed on the $r$-band SDSS FITS images for barred galaxies. Using $ra$ and $dec$ information of each galaxy, we set the center of ellipses and find parameters along ellipse annuli. Typical barred galaxies show maximum ellipticity at the end of the bar, and position angles are uniform within the bar structure. We double-checked visually-measured bar lengths by using these constraints. In Figure~\ref{leng}, we compare bar lengths which are determined from using two different methods: visual measurement and ellipse fitting. The agreement is good. We use the bar lengths determined through visual measurement because the ELLIPSE task occasionally fails to converge to a solution while visual measurement works even in complicated cases.

\subsection{Spectral line data and velocity dispersion}
The SDSS spectral database covers the wavelength range of 3800 -- 9200 \AA. Oh et al. (2011) recently released new and improved line measurements on the SDSS galaxies using GANDALF routine (Gas AND Absorption Line Fitting; Sarzi et al. 2006). Oh et al. (2011) measured the stellar kinematics using the publicly-available penalized pixel-fitting code (pPXF; Cappellari \& Emsellem 2004). Readers are referred to Oh et al. (2011) for details.

GANDALF also made estimates of the central velocity dispersion, $\sigma_{\rm ap}$. Its improved algorithm allows for accurate measurement even for galaxies with low (and thus difficult-to-detect) velocity dispersions. SDSS provides velocity dispersion measurements only for 33\% of our sample galaxies, failing for the rest. Oh et al. (2011)'s technique succeeds in measuring velocity dispersions for most of our galaxies. We have however decided to accept only the velocity dispersions that are greater than their associated error. 5,255 out of 6,658 (that is, 78\%) satisfy this condition. We used the following formula of Cappellari et al. (2006) for aperture correction of the velocity dispersion:
\begin{eqnarray}
\sigma_{\rm e} = (\,\frac{R_{\rm ap}}{R_{\rm e}}\,)^{0.066\,\pm\,0.035} \;\sigma_{\rm ap},
\end{eqnarray}
where $R_{\rm e}$ is the effective radius of the galaxy and $R_{\rm ap}$ is the aperture radius of the SDSS fiber (1.5 $''$). The effective radius given by a de Vaucouleurs fit which follows the light of the bulge component, $R_{\rm dev}$, is converted to an effective circular radius, $R_{\rm e}$, using isophotal axes in the SDSS $r$-band (Bernardi et al. 2003):
\begin{eqnarray}
R_{\rm e} = \sqrt{IsoB_{\rm r}/IsoA_{\rm r}}\; R_{\rm dev}.
\end{eqnarray}
Note that this formula is mainly for early-type galaxies and thus may cause a substantial uncertainty in measuring the effective velocity dispersion.

\begin{figure}[t]
\centering 
\includegraphics[width=0.45\textwidth]{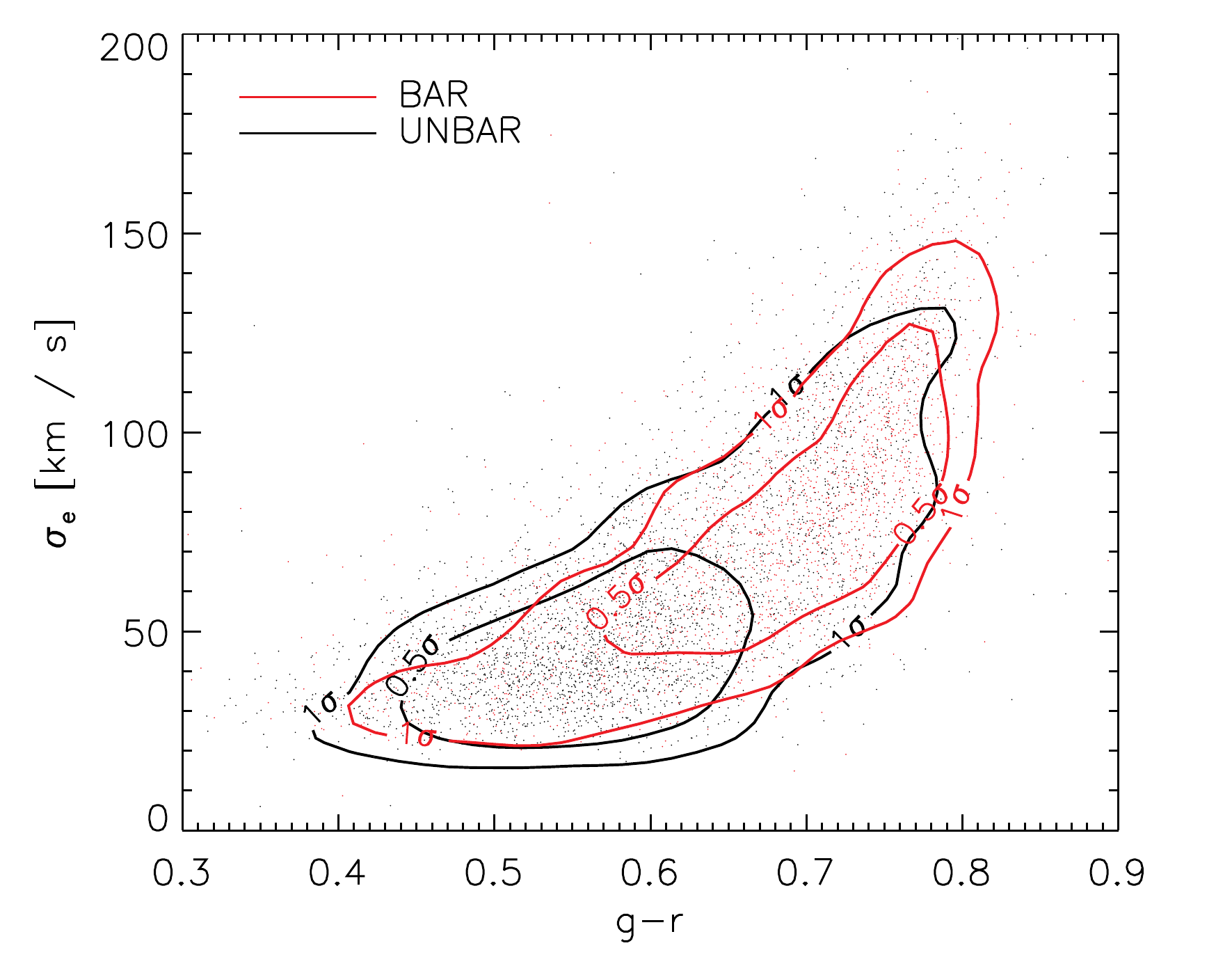}

\caption{Stellar velocity dispersion of our sample as a function of $g-r$ color. For comparison, the 0.5-$\sigma$ and 1-$\sigma$ contours of barred and unbarred galaxies are also shown.}
\label{sig}
\end{figure}

Figure~\ref{sig} shows effective velocity dispersion for our sample. Despite similar overall ranges of velocity dispersion in barred and unbarred galaxies, barred galaxies tend to have higher velocity dispersion, as indicated by the 0.5-$\sigma$ contour. Stellar velocity dispersion is often considered as an indicator to supermassive black-hole mass and galaxy dynamical mass.
 
\subsection{Emission-line diagnostics}
In order to study bar effects on central star formation and AGN activity, we first attempt to classify our galaxies into star-formation dominant and AGN-dominant galaxies. We use the ``BPT diagnostics'' (Baldwin, Phillips \&\ Terlevich 1981) which are based on emission flux ratios between Balmer and forbidden lines (e.g., Kauffmann et al. 2003). In this analysis we use $[OIII]\;\lambda \,5007/H\beta \;\lambda\, 4861$ and $[NII]\; \lambda\; 6583/H\alpha\; \lambda \;6563$ emission flux ratios. We apply GANDALF's parameter ``amplitude-over-noise'' (A/N) to select galaxies with strong emission lines. A/N represents the strength of each {\em emission} line compared to the noise level of the nearby continuum. Galaxies with A/N above 3.0 in all four lines are classified as ``significant-emission'' galaxies. Of our 6,658 galaxies, 59\% are classified as ``significant-emission'' galaxies.

\begin{table}[t]
 \begin{center}
  \caption[Results of spectral line classification]
  {Results of spectral line classification}
  \begin{tabular}{lccc}
  \hline \hline
 \multicolumn{1}{c}{Classification} &Total (6658) & Bar (2422) & Unbar (4236) \\
 \hline
  Weak Emission & 40.9\% (2724) &31.6\% (765) & 46.3\% (1959) \\
 Strong Emission & 59.1\% (3934) &68.4\% (1657) & 53.8\% (2277) \\
 \hline
\quad Star formation & 38.1\% (2537) &38.9\%  (942) & 37.7\% (1595) \\
\quad AGN &  21.0\% (1397) & 29.5\%  (715) & 16.1\% (682) \\
   \hline \hline \\
\end{tabular}
\label{tab:bpt}
\end{center}
\end{table}

\begin{figure}[t]
\centering
\includegraphics[width=0.45\textwidth]{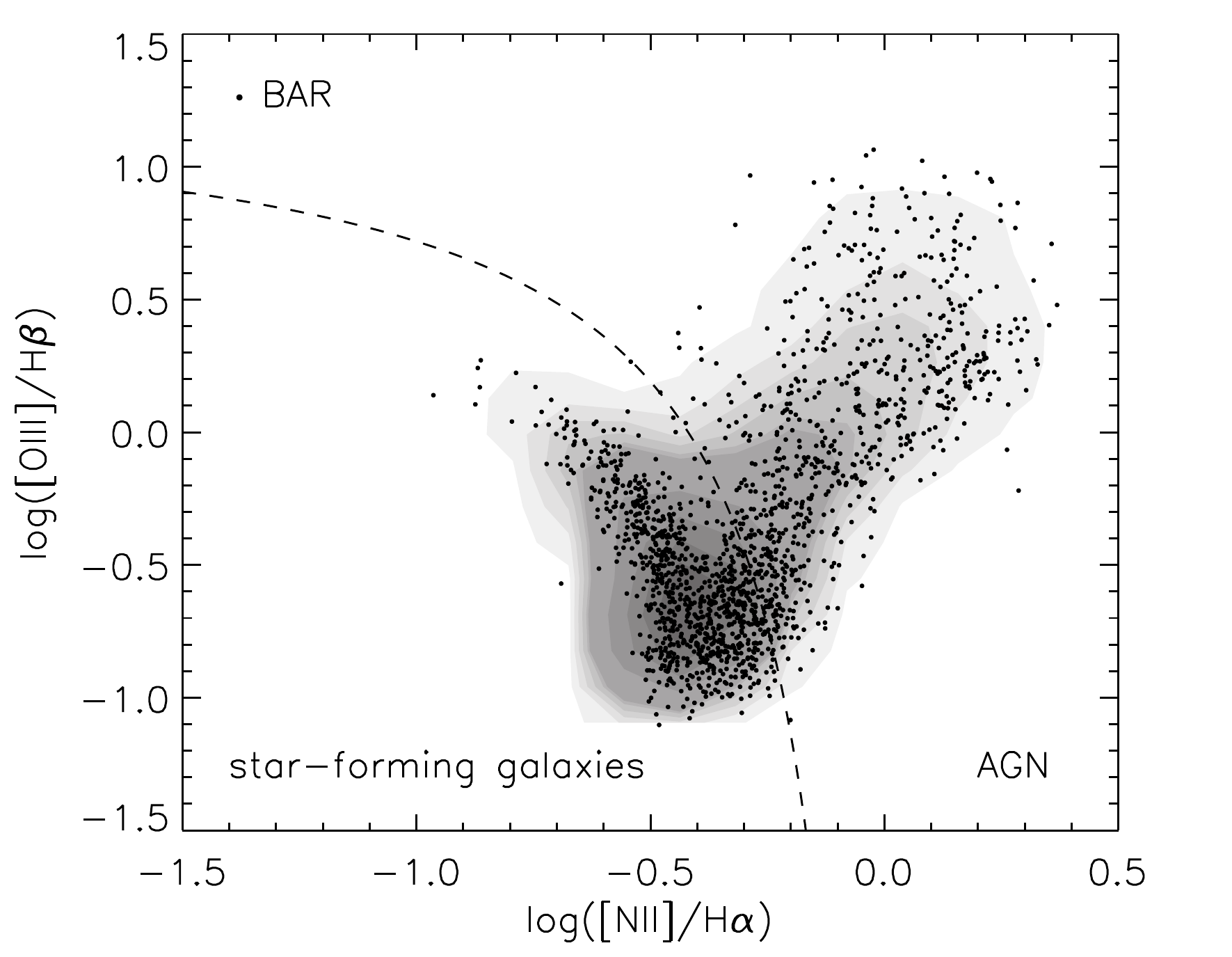}

\caption{The BPT diagram, as an AGN diagnostic (Baldwin, Phillips \&\ Terlevich 1981). Only significant emission galaxies (A/N $>$ 3 for all four lines) are shown in the figure. Black dots and the shaded contours show the distribution of barred and unbarred galaxies, respectively. The dashed line presents the demarkation line from Kauffmann et al. (2003) which divides star-forming galaxies and AGN.}
\label{bpt}
\end{figure}

Figure~\ref{bpt} shows the distribution of our galaxy sample on a BPT diagram. Star-forming galaxies and AGN have been separated using the empirical demarkation line defined by Kauffmann et al. (2003). Only ``significant-emission'' galaxies are shown in the figure. Barred galaxies show a higher fraction of ``significant-emission'' galaxies (68\%) compared with unbarred galaxies (53\%). Star-forming galaxies are more common in both barred and unbarred galaxies but much more dramatically among unbarred galaxies. The details are listed in Table~\ref{tab:bpt}. However, such a simplistic statistical analysis is insufficient to study the bar effects because bar effects are degenerate with underlying relations between galaxy properties. We will discuss this in Section 4.

\subsection{Stellar mass}
\begin{figure}[t]
\centering
\includegraphics[width=0.45\textwidth]{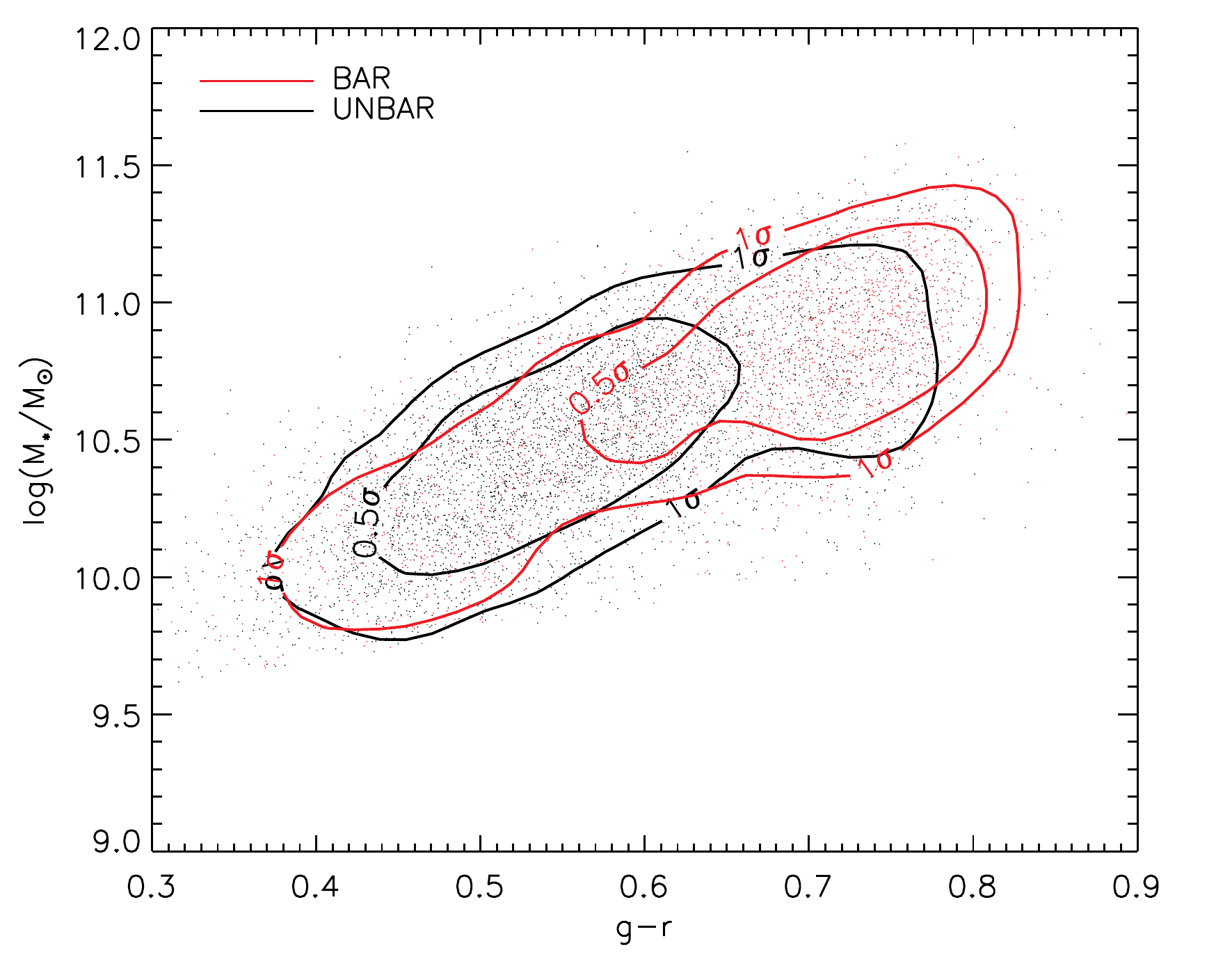}

\caption{Stellar mass of our sample galaxies as a function of $g-r$ color. The 0.5-$\sigma$ and 1-$\sigma$ contours of barred and unbarred galaxies are shown for comparison.}
\label{ms}
\end{figure}

The stellar mass of each galaxy is calculated from its colors and luminosities using the following formula from Bell et al. (2003):
\begin{eqnarray}
log\left(\frac{M_*}{M_\odot}\right)=-0.306+1.097(g-r)-0.4(M_{\rm r}-M_{\rm r,{\odot}}).
\end{eqnarray}
Figure~\ref{ms} shows the estimated stellar masses of our sample. From its contours we notice different distributions of stellar mass in barred and unbarred galaxies. We find that barred galaxies are relatively more massive than unbarred galaxies.

\subsection{Black-hole mass}
The presence of supermassive black holes in elliptical galaxies and spiral bulges is widely accepted (e.g., Kormendy \& Richstone 1995 and references therein). A realistic attempt to estimate black-hole masses began after a correlation was found between galactic black-hole mass and bulge luminosity (e.g., Dressler 1989; Magorrian et al. 1998). A tighter correlation has been found between central black-hole mass and the line-of-sight velocity dispersion (e.g., Gebhardt et al. 2000). So, we adopt M$_{\rm BH}$-$\sigma$ relation to estimate black-hole mass. 
The general form of the M$_{\rm BH}$-$\sigma$ relation is given by
\begin{eqnarray}
   log({M_{\rm BH}}/{M_\odot}) = \alpha + \beta\,log({\sigma_{\rm e}}/200\,km\,s^{-1}) .
\end{eqnarray}
We adopt the $M_{\rm BH}$-$\sigma$ relation with ($\alpha$, $\beta$) = (7.67 $\pm$ 0.115, 4.08 $\pm$ 0.751) for barred galaxies and ($\alpha$, $\beta$) = (8.19 $\pm$ 0.087, 4.21 $\pm$ 0.446) for unbarred galaxies from G$\rm \ddot{u}$ltekin et al. (2009). The reason for different $M_{\rm BH}$-$\sigma$ relations is that the stellar motion along a bar effectively enhances the ``measurable'' central velocity dispersion. 




\section{Results}
\subsection{Color-magnitude relation}
Galaxies are populated in three main areas of the color-magnitude diagram (CMD): ``the red sequence'', ``the green valley'', and ``the blue cloud'' (Strateva et al. 2001, Bell 2003). Generally speaking, early-type galaxies form a narrow red sequence (e.g. Bower, Lucey \& Ellis 1992; Ellis et al. 1997; van Dokkum et al. 2000), and late-type galaxies reside in the blue cloud area. Green-valley galaxies, located between the red sequence and the blue cloud, show a mixture of galaxy morphologies.

\begin{figure}[t]
\centering 
\includegraphics[width=0.45\textwidth]{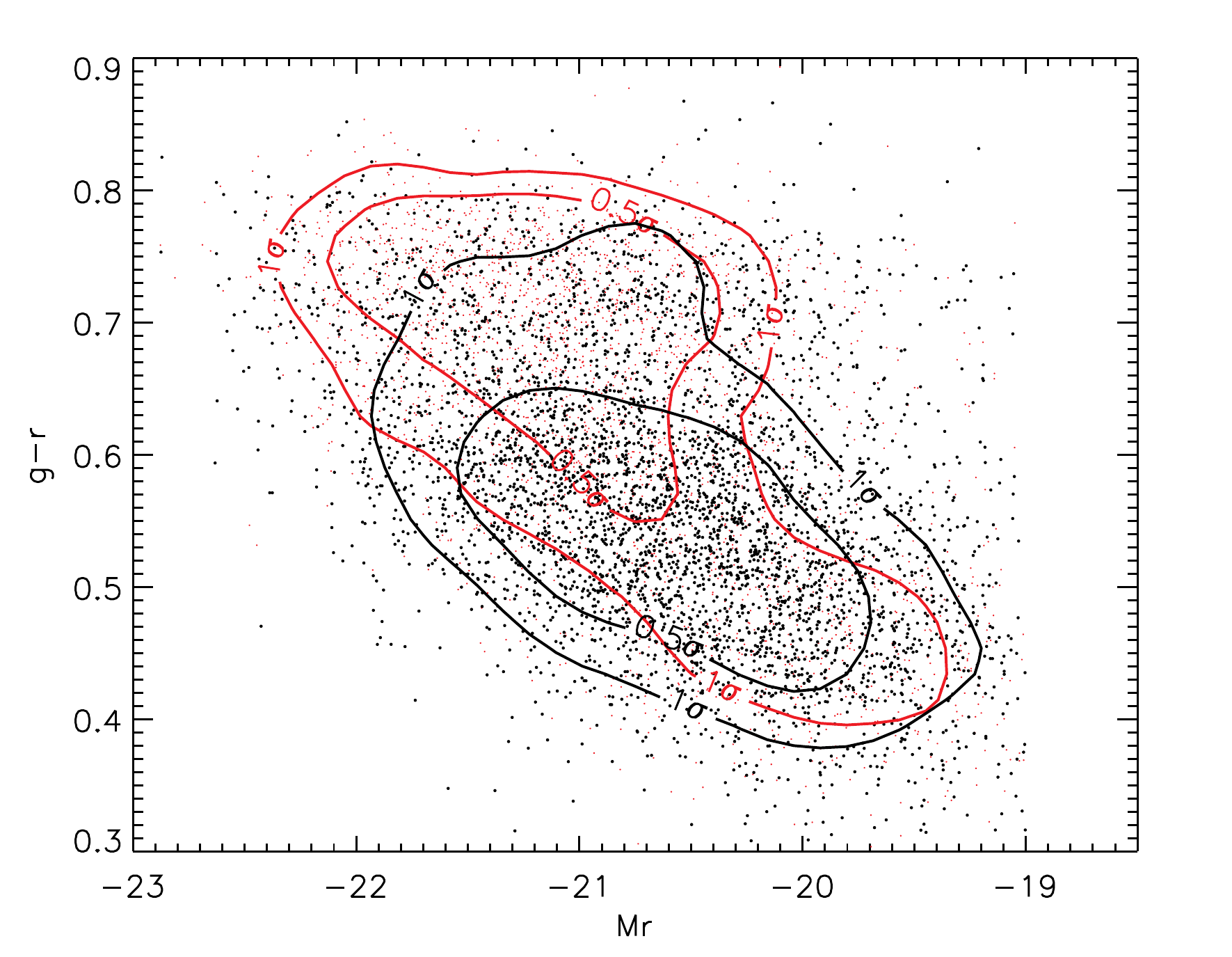}

\caption{Color-magnitude diagram for our late-type galaxies. The 0.5-$\sigma$ and 1-$\sigma$ contours of barred and unbarred galaxies are also shown for comparison.}
\label{cmd1}
\end{figure}

The CMD based on $g-r$ color is plotted in Figure~\ref{cmd1}. 
Our morphological criteria for selecting late-type galaxies allow a wide baseline in galaxy color. Our late-type sample contains not only typical late-type galaxies in the blue cloud but also a number of red spirals in the green valley. From the optical CMD, we have found that barred galaxies occupy different areas in the CMD from unbarred galaxies. The majority of unbarred galaxies are located in the blue cloud. A significant number of barred galaxies, by contrast, are redder and brighter than typical late-type galaxies, so they are shown in the green valley. Though we do not display here because it feels redundant, we also found that barred galaxies have a substantially higher mean value of concentration index in our exercise. Hence, it is essential {\em not} to have any color bias (cut) for the study of bar effects. This is consistent with the findings in the previous sections.

\begin{figure}
\centering 
\includegraphics[width=0.45\textwidth]{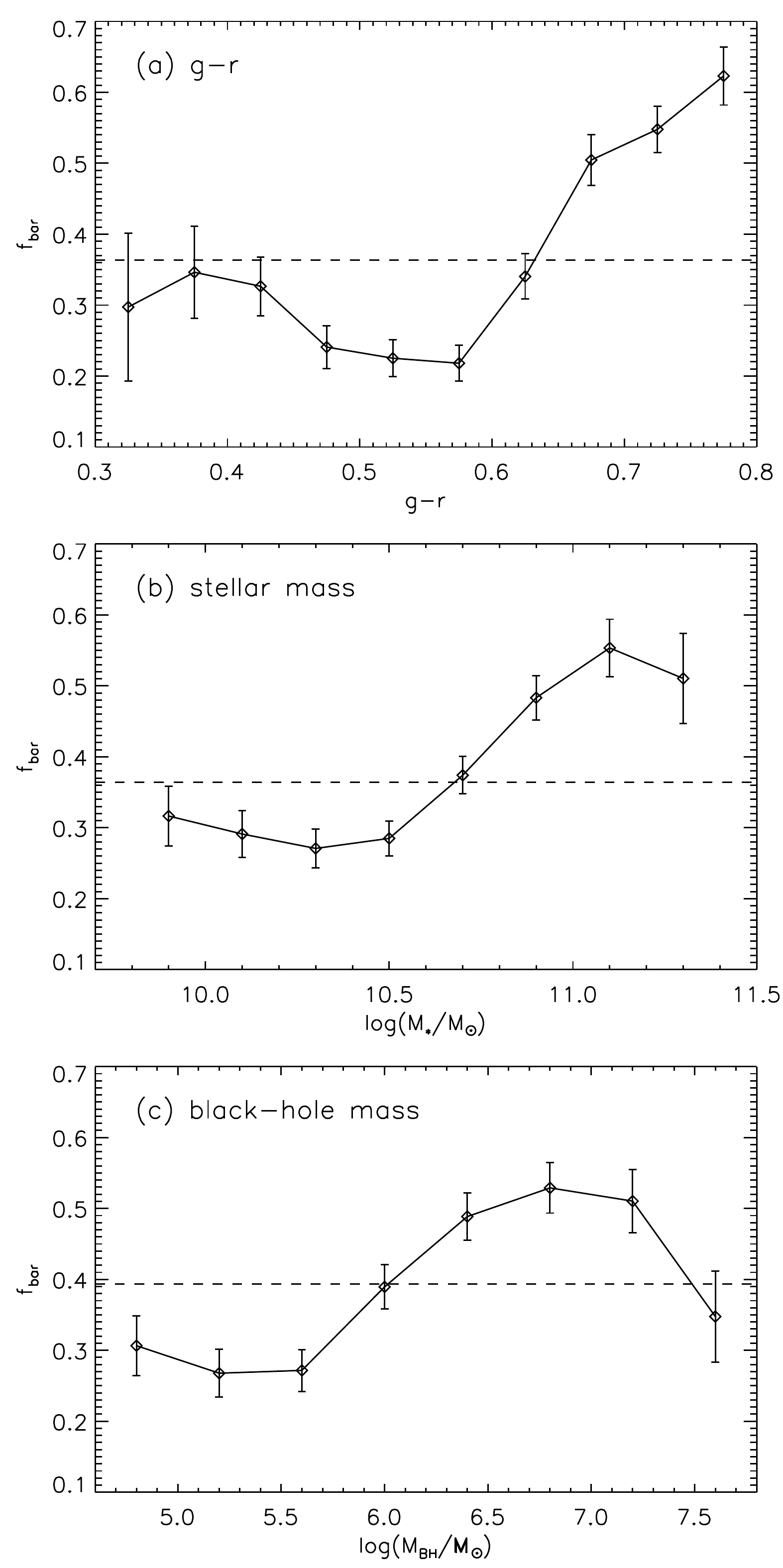}
\caption{Bar fractions against galaxy properties. The dashed line shows the mean bar fraction for our sample. Poisson errors are denoted as error bars. The bar fractions are plotted as a function of (a) $g-r$ color, (b) stellar mass, and (c) black-hole mass. The value of $f_{\rm bar}$ is slightly different for (c) because we have velocity dispersion measurements only for 78\% of our sample, as mentioned in Section 3.2. The bar fraction shows the non-monotonic increase for galaxy parameters.}
\label{fbar}
\end{figure}

\subsection{Bar fraction}
We investigated the bar fraction which is measured from the SDSS color-composite images. We found 2,422 barred galaxies out of 6,658 late-type galaxies, yielding a bar fraction of 36$\%$. Visual detection is biased toward stronger bars, and weaker bars are likely to be missed more often. According to the studies using the RC3 catalogue, the bar fraction for strong bars (SB types) is around 30\%; but if weak bars (SAB types) are included the bar fraction increases to about 60\% (the RC3; de Vaucouleurs et al. 1991; Ho et al. 1997; Laurikainen \& Salo 2004). Our sample and result appear to be more allied to classical strong bars. Masters et al. (2011), who visually selected barred galaxies from SDSS (through the Galaxy Zoo ``citizen'' effort), found similar results to ours. Their mean bar fraction is 29.4\% from the sample of 13,665 disk galaxies. Note that their sample is twice the size of ours, as we have imposed stricter sampling criteria (e.g., inclination).

Recent studies based on the ellipse fitting technique for selecting barred galaxies from SDSS reported 45 - 50\% of optical bar fractions (Barazza et al. 2008; Aguerri, M$\rm \acute{e}$ndez-Abreu, \& Corsini 2009). Such a high fraction compared to ours (36\%) could be due to either the difference in the sample selection (visual classification vs. parameter cut) or the bar classification method (visual classification vs. ellipse fitting). Barazza et al. (2008) applied a color cut  and Aguerri, M$\rm \acute{e}$ndez-Abreu, \& Corsini (2009) used concentration index to classify galaxy morphology. So, red spiral galaxies are excluded from their samples. However, if differences in sample selection were the main reason, their bar fraction should be lower than ours, because we found a lower bar fraction on blue galaxies. Moreover, Aguerri, M$\rm \acute{e}$ndez-Abreu, \& Corsini (2009), who did both morphological classification and ellipse fitting to select barred galaxies, found higher bar fraction when they use ellipse fitting method (45\%) than visual classification (38\%). Therefore the primary reason for the difference must be the bar classification method. Ellipse fitting may have the advantage of detecting weak barred galaxies. In this sense, one might think that visual detection of bars places a lower limit for the bar fraction. 

We present the bar fractions of our sample {\em as a function of galaxy properties} in Figure~\ref{fbar}. It seems that the bar fraction is higher on galaxies having earlier morphology: we find that more massive galaxies have higher bar fractions when we analyze the bar fraction according to either stellar mass or black-hole mass. This result is consistent with recent studies. Sheth et al. (2008) showed variations of the bar fraction with respect to galaxy properties, such as galaxy color, mass, and luminosity. Giordano et al. (2010) found a bar fraction of around 50\% in ``early'' disk galaxies but around 25\% for ``late'' disk galaxies. Nair \& Abraham (2010) also suggest that the bar fraction can be highly affected by galaxy morphology. 

The bar fractions behave non-monotonically in galaxy properties. The bar fraction slightly decreases with $g-r$ until $g-r$ becomes 0.6. However, it rapidly bounces back to $f_{\rm bar} \sim 0.6$ as $g-r$ gets redder. Nair \& Abraham (2010) found a similar result for stellar mass. In fact, such non-monotonic features in the bar fraction have been reported earlier using the RC3 catalogue. Odewahn (1996) and Elmegreen, Elmegreen, \& Hirst (2004) compared bar fractions according to galaxy morphology and found that the bar fraction (considering only strong bars) hits the minimum in Sc types but increases in for both earlier and later types. Moreover, this is supported by the bar fractions measured by Masters et al. (2011), which is also based on morphological classification. The fractions of Masters et al. (2011) are lower than ours by 7\%, but the overall trend of bar fraction with respect to $g-r$ is remarkably consistent with ours (see Figure 3 of Masters et al. 2011). Barazza et al. (2008) did not recover the rising trend for redder galaxies simply because their sample was based on a color cut and thus missed red spirals. 

\begin{figure}
\centering
\includegraphics[width=0.45\textwidth]{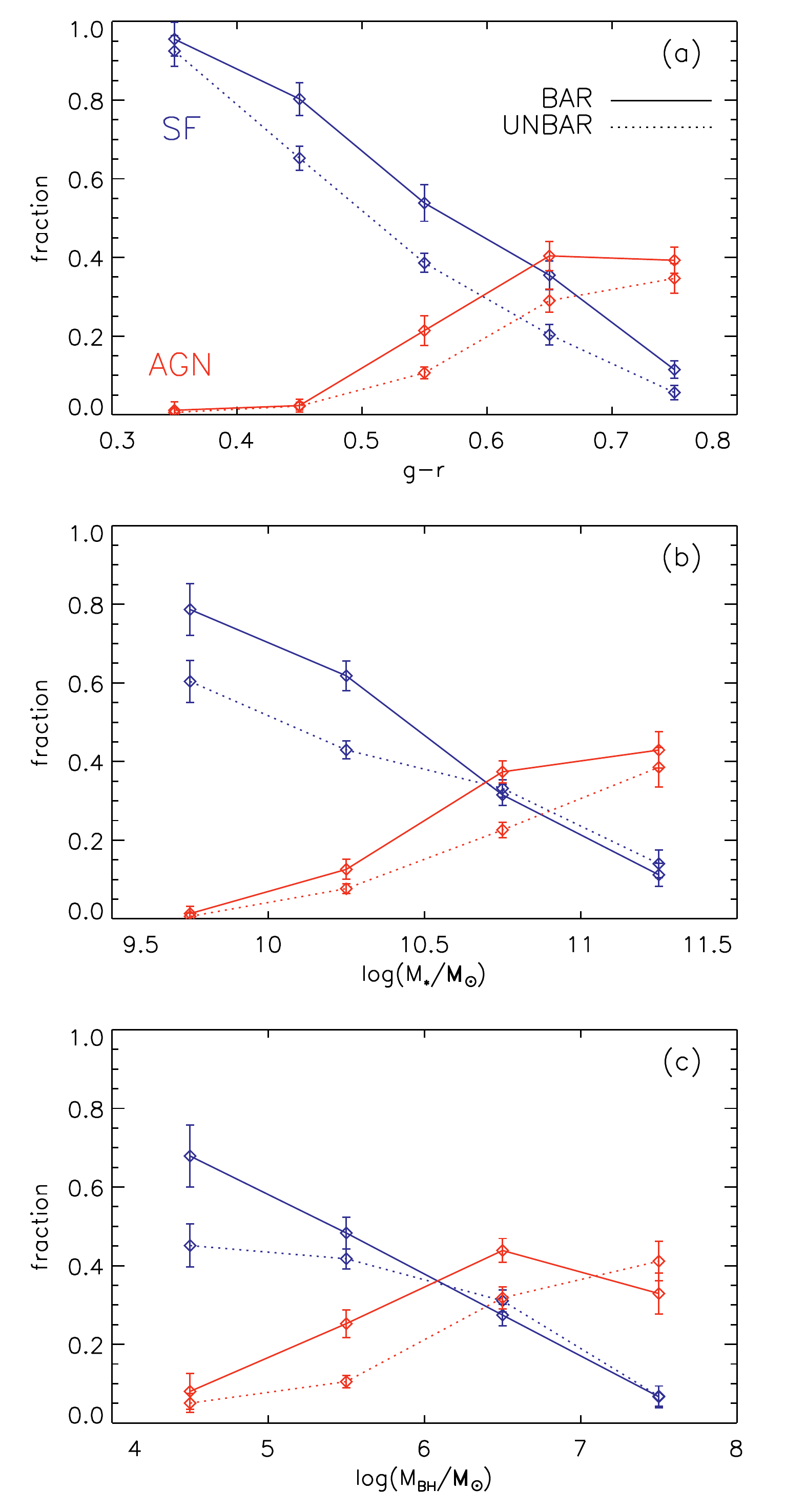}
\caption{Galaxy fractions showing central star formation and AGN activity with respect to galaxy properties. The blue and red lines represent central star formation and AGN, while the solid and dotted lines indicate barred and unbarred galaxies, respectively. It is shown as a function of (a) $g-r$, (b) stellar mass, and (c) black-hole mass. Poisson errors are shown by error bars. }
\label{frac1}
\end{figure}

\begin{figure*}
\centering
\includegraphics[width=1\textwidth]{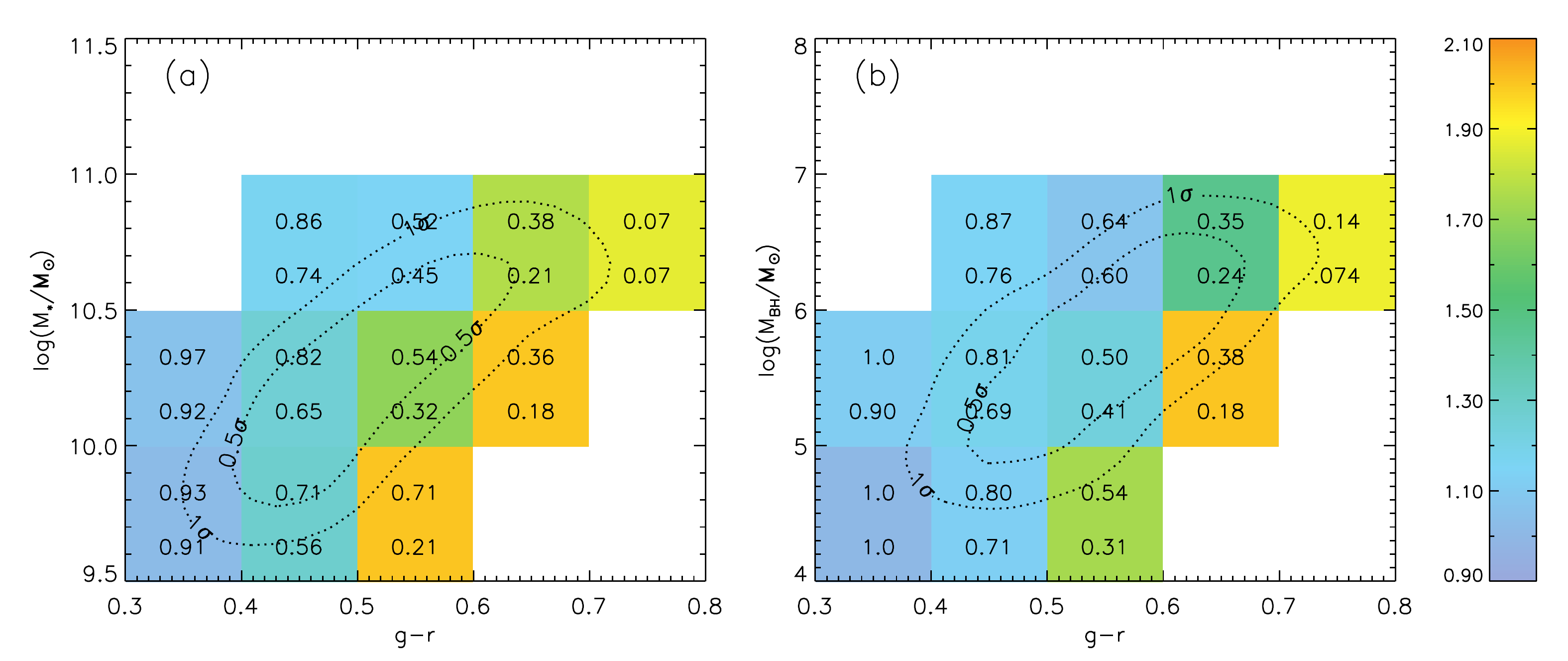}
\caption{The stellar (a) and black hole (b) mass-color grid diagram for star-forming galaxies. 
Each grid has two numbers, indicating the fraction of star-forming activity in barred (top) and unbarred (bottom) respectively. The color key on the right indicates $f_{\rm norm}$ which shows the bar effect in the number of galaxies with central star formation or AGN activity. Dotted lines indicate the 0.5-$\sigma$ and 1-$\sigma$ contours for galaxies with star-formation activity.}
\label{grid1}
\end{figure*}

\begin{figure*}
\centering
\includegraphics[width=1\textwidth]{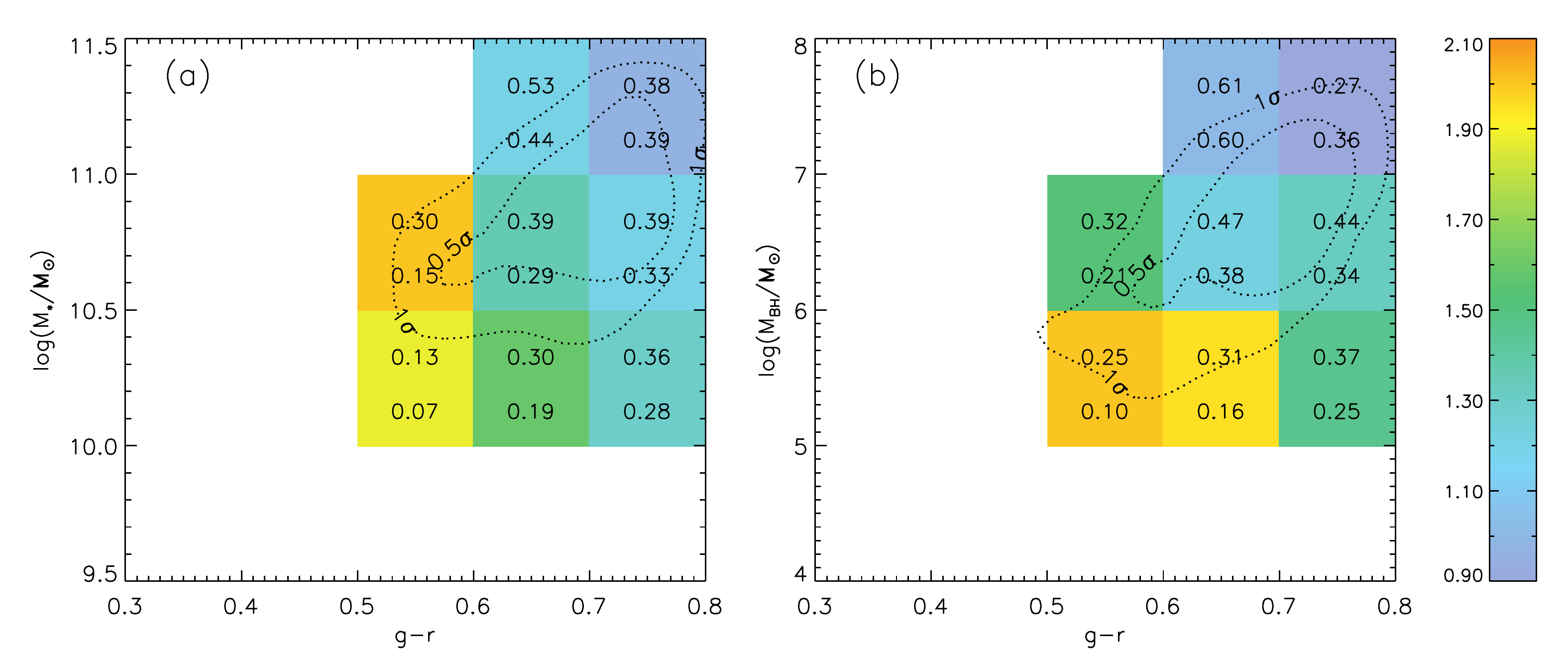}
\caption{
Same as Figure~\ref{grid1}, but for AGN.}
\label{grid2}
\end{figure*}

\begin{figure*}
\centering
\includegraphics[width=1\textwidth]{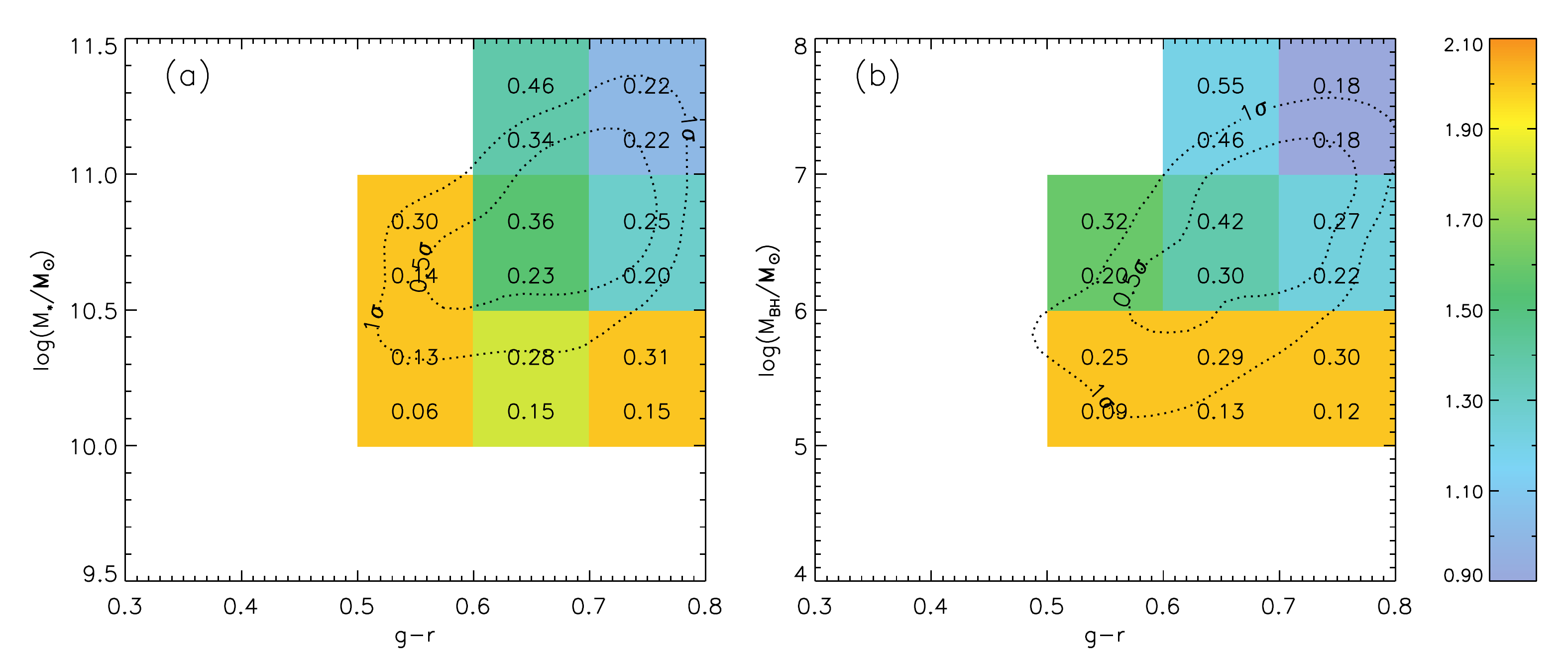}
\caption{
Same as Figure~\ref{grid2}, but for AGN with higher $W(H\alpha)$ than 3\,\AA.}
\label{grid22}
\end{figure*}

\subsection{Bar Effects on Fractions of central star formation and AGN activity}\label{baronnuc}
In this study, we directly compare the fractions of galaxies showing specific star formation and AGN activity in barred and unbarred galaxies. The fractions of galaxies with (significant) central star formation in barred and unbarred galaxies are similar, at 38.8\% (barred) and 37.4\% (unbarred). The fraction of AGN in barred galaxies (29.5\%) on the other hand is substantially higher than that of unbarred galaxies (16.2\%). However, bar galaxy fractions are highly affected by galaxy properties as shown in Figure~\ref{fbar} and so are central star formation and AGN activity. Consequently, the comparison of ``total'' fractions of galaxies with central star formation or AGN activity between barred and unbarred galaxies is inappropriate for showing the whole nature of bar effects.
 
In Figure~\ref{frac1}, we plot the fractions of galaxies which show central star formation and AGN activity for barred and unbarred galaxies. We have inspected the overall trends with respect to optical colors, stellar mass, and black-hole mass and found clear dependencies on galaxy properties. The fraction of galaxies with central star formation dramatically decreases with galaxy optical color (panel a), stellar mass (panel b), and black-hole mass (panel c). On the contrary, AGN are more frequent in redder and/or more massive galaxies.

Bar effects on AGN fractions appear to be significant only in intermediate-property galaxies. On the other hand, bar effects on central star formation are clear throughout the whole color range (panel a) but only in the low-mass range (panel b and c). One may wonder why bar effects on central star formation are clear in red (e.g., $g-r \approx 0.65$) galaxies but not in massive (e.g., $log M/M_\odot \approx 10.7$) ones considering the general correlations between optical colors and stellar/black-hole mass. To answer this question, we have devised grid diagrams which illustrate bar effects in the two-dimensional parameter space. We define a new index $f_{\rm norm}$ which quantitates bar effects, as follows,
\begin{eqnarray}
f_{\rm norm} = \frac{f_{\rm activity,bar}}{f_{\rm activity,unbar}},
\end{eqnarray}
where $f_{\rm activity,bar/unbar}$ is the fraction of each activity for barred/unbarred galaxies. An $f_{\rm norm}$ greater than 1.0 indicates a positive bar effect.

Figure~\ref{grid1} shows the grid diagram of color against stellar/black-hole mass. In each pixel, the upper and lower values are the galaxy fractions with (significant) central star formation in barred and unbarred galaxies, respectively. The dotted lines show 0.5-$\sigma$ and 1.0-$\sigma$ contours for the galaxy distribution. The color of each grid indicates $f_{\rm norm}$. It is apparent that the color dependence (horizontal sequence) is clearer than the stellar/black-hole mass dependence both in the monotonic trend and in amplitude (numerals in each pixel). This is probably why bar effects are more clearly visible in the whole range of color, whereas it is not the case with stellar/black-hole mass in Figure~\ref{frac1}.

For a fixed stellar/black-hole mass, bar effects are stronger in redder galaxies. For fixed colors, mass effects are visible but less clear. 

The grid diagrams for AGN are shown in Figure~\ref{grid2}. The distribution of AGN in these figures suggests that late-type AGN are more massive and redder compared with galaxies showing central star formation. Further, they are present in a smaller area on the color-mass plane. Color and mass trends are all less clear compared with the galaxies with central star formation; this was already visible in Figure~\ref{frac1}. Our results suggest that {\em the fractions of central star formation and AGN activity are clearly higher in barred galaxies than in unbarred galaxies}; but bar effects are not always clearly visible if degenerate correlations between bar effects and galaxy properties are not properly broken.

Our BPT-based AGN classification allows a fair number of LINERs and might be contaminated by LINER-like emission powered by old stars (Sarzi et al. 2010; Cid Fernandes et al. 2011). According to the classification established by Cid Fernandes et al. (2011) where the equivalent width of Ha of true AGN exceeds 3\,\AA, 23\% of our AGN are classified as ÒAGN-likesÓ. Excluding them from our analysis, we tested our main results (Figure~\ref{grid22}), and found that the change of the AGN criterion does not affect the overall conclusion. We adopt the BPT diagnostics throughout this study. But it will be appropriate to consider the new stricter classifications when confirmed through more definite tests.

\subsection{Bar Effects on Emission line strengths}
We compared H$\alpha$, [NII], and [OIII] emission line luminosities (hereafter ``emission luminosities'') of barred and unbarred galaxies. The H$\beta$ emission line strength is determined from H$\alpha$ based on the concept of Balmer decrement, so we exclude H$\beta$ in this analysis. 
Emission luminosities naturally correlate with the stellar mass of galaxies. So, direct comparison of line strength can be biased to this obvious mass trend. To show the difference in emission luminosity between barred and unbarred galaxies free from the mass-luminosity relation, we use Òspecific emission luminositiesÓ defined as the emission luminosities divided by the fiber luminosity in the $r$-band. Fiber flux for the SDSS 3$''$ aperture is provided by the SDSS database. 

\begin{figure}
\centering
\includegraphics[width=0.45\textwidth]{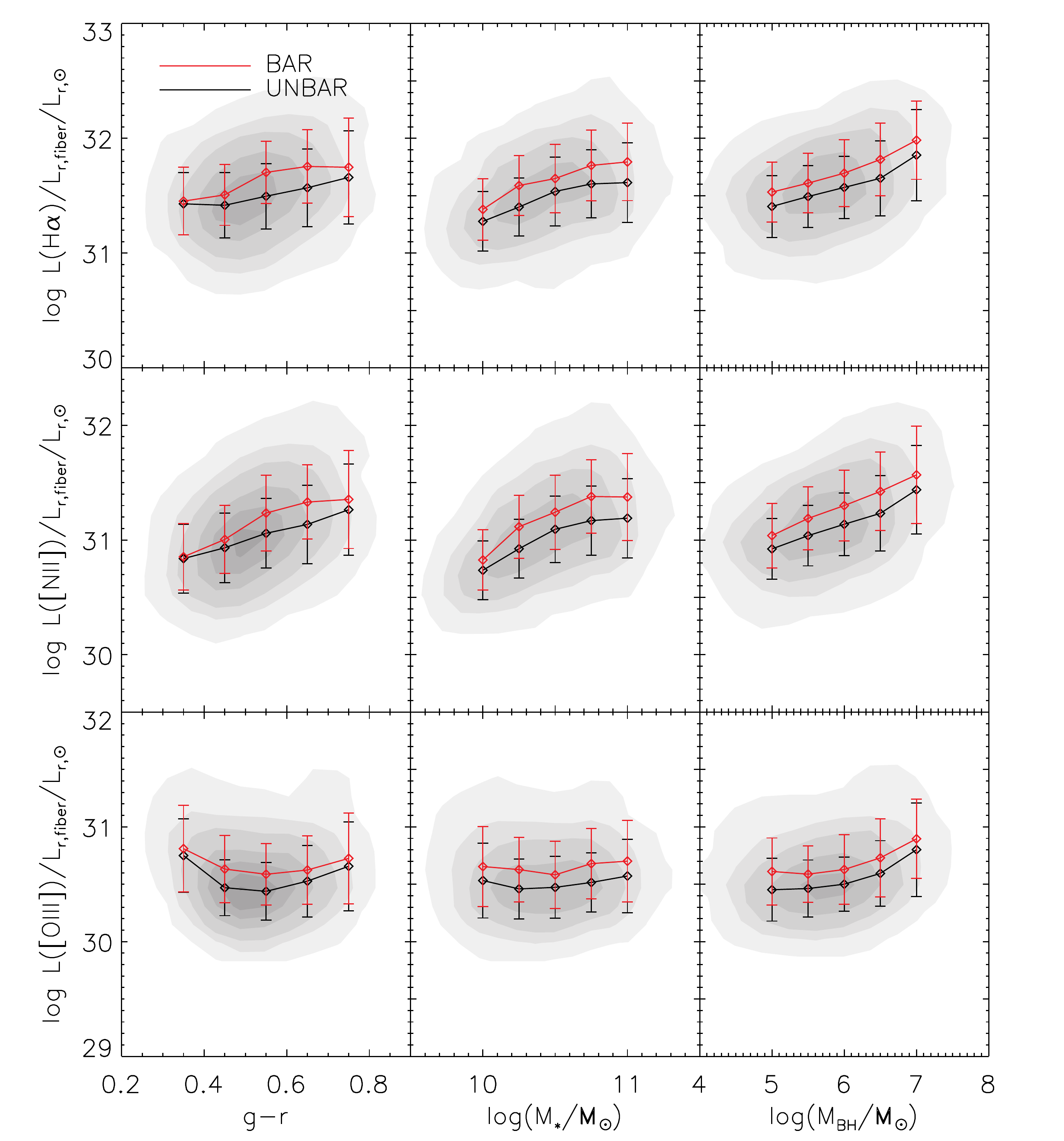}
\caption{Specific emission luminosities against galaxy properties for star-forming galaxies.
The shaded contours show the total distribution of our sample, and red and black points show the {\em mean} specific luminosity for barred and unbarred galaxies, respectively. The error bars show 68\% probability distributions for each bin.}
\label{SFflux1}
\end{figure}

\begin{figure}
\centering
\includegraphics[width=0.45\textwidth]{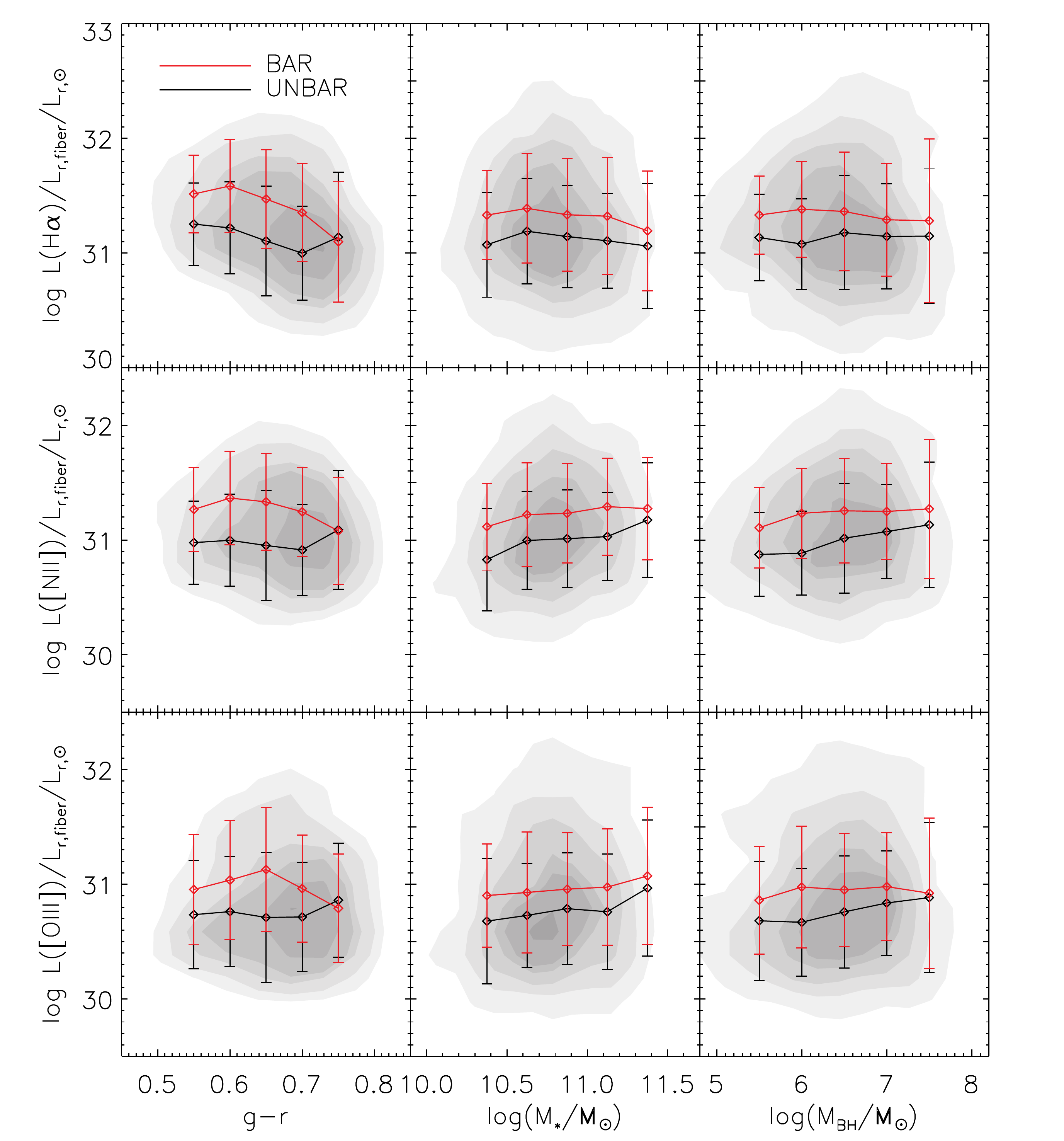}
\caption{Same as Figure \ref{SFflux1}, but for AGN.}
\label{AGNflux1}
\end{figure}

Figure~\ref{SFflux1} shows the specific H$\alpha$, [OIII], and [NII] emission luminosities against $g-r$, stellar mass, and black-hole mass of star-forming galaxies. Mean specific luminosities for each emission line with respect to galaxy properties (diamonds) and 68\% probability distributions in luminosity (error bars) are displayed. The first thing to notice is that H$\alpha$ and [NII] show a positive, although weak, correlation with galaxy properties. This looks counter-intuitive, because redder galaxies are generally less active in terms of star formation. In fact, our data are also consistent with this expectation (Figure~\ref{frac1}). Combining these two empirical facts, we may conclude that redder and more massive late-type galaxies are less likely to host (significant) central star formation, but once they become significantly active their emission strengths become strong for their stellar/black-hole mass. This would imply that central star formation is more bursty in redder late types. 

Bar effects on the (specific) emission strengths are small but ubiquitous in our parameter space, which probably means that they are statistically significant. This is consistent with the earlier work of Ellison et al. (2011) which found higher SFR in barred galaxies compared to unbarred galaxies when stellar mass of galaxies are higher than $10^{10}$ M$_\odot$.

AGN display a slightly different result (Figure \ref{AGNflux1}). The trends against galaxy properties are not visible. On the other hand, bar effects are even stronger than in the case of star formation. If we may interpret the emission luminosities as AGN strengths, these results would imply that AGN strength is enhanced by the presence of a bar and linearly correlates with stellar or black-hole mass. Considering the Magorrian relation (Magorrian et al. 1998), finding the same trend against stellar mass and black-hole mass is sensible. 

In conclusion, bar effects are clearly visible in the emission line strengths, too.

\begin{figure}
\centering
\includegraphics[width=0.45\textwidth]{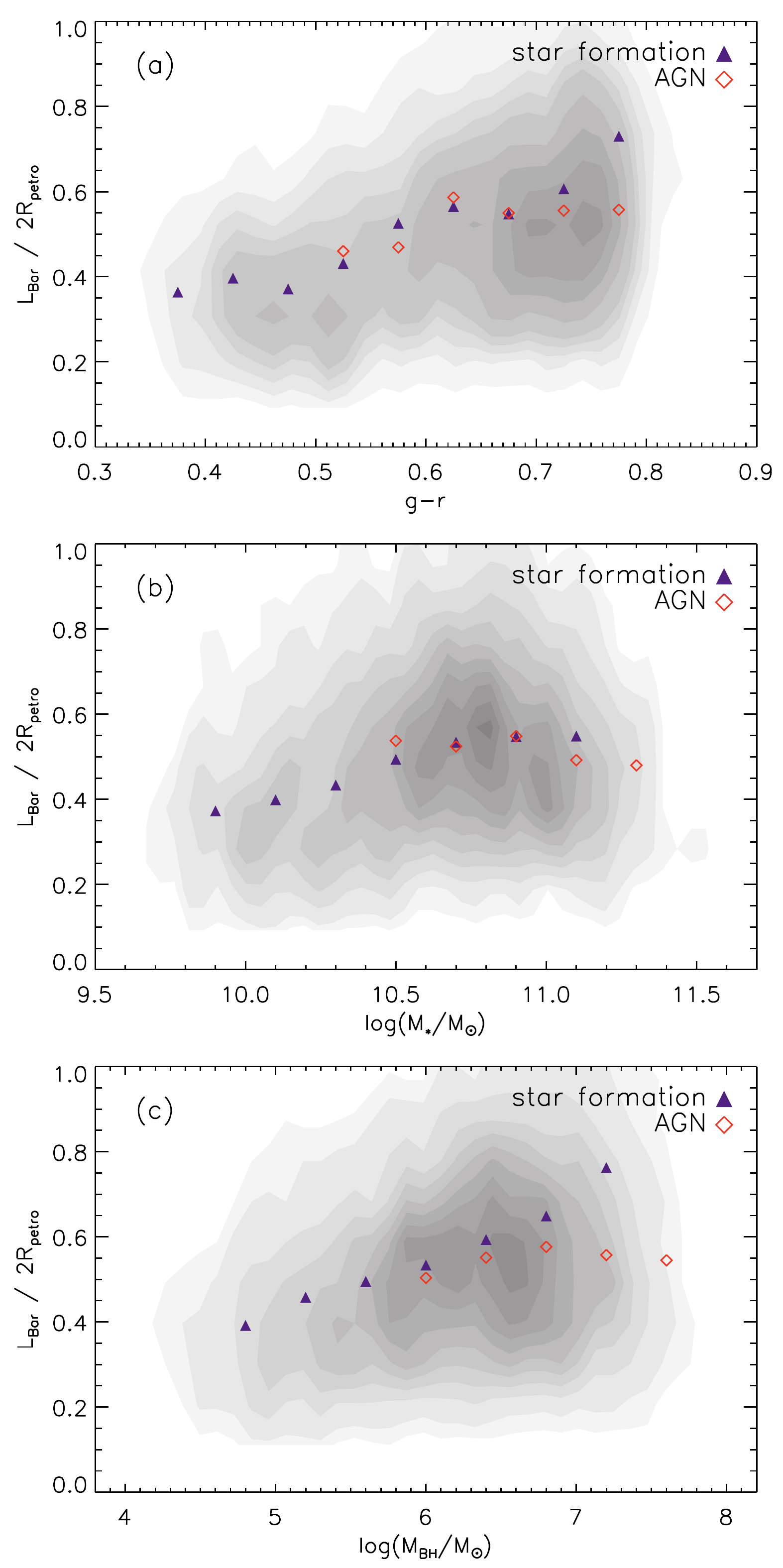}
\caption{Relative bar length with respect to galaxy properties. Relative bar lengths are plotted as a function of (a) $g-r$ color, (b) stellar mass, and (c) black-hole mass. Blue and red symbols represent the median values of relative bar lengths of star formation and AGN, respectively. The median value of relative bar lengths increases according to the galaxy color and stellar/black-hole mass, but star-forming galaxies and AGN show different trends.}
\label{bleng1}
\end{figure}

\subsection{Bar lengths}
Bar length is normalized to twice the $r$-band petrosian radius of the galaxy to negate the distance effect on the apparent bar length. We call this ``relative bar length''. It is known that petrosian radius is a good proxy for the optical light of a galaxy.

In Figure \ref{bleng1}, we plot relative bar length as a function of $g-r$ color, stellar mass, and black-hole mass to investigate the correlation between bar length and galaxy morphology. We found that relative bar length generally increases with $g-r$ color and stellar/black-hole mass, which is in agreement with theoretical expectations (e.g., Athanassoula \& Martinet 1980) and previous observational studies (Martin 1995; Laurikainen, Salo, \& Buta 2004; Erwin 2005; Elmegreen et al. 2007; Hoyle et al. 2011; Gadotti 2011). Interestingly, when we break the sample into star-forming and AGN galaxies, it is {\em only} star-forming galaxies that show a tight correlation. The difference between star-forming and AGN galaxies for given morphology (color or mass) in relative bar length implies that they are in different stages of bar evolution.

\begin{figure}
\centering
\includegraphics[width=0.45\textwidth]{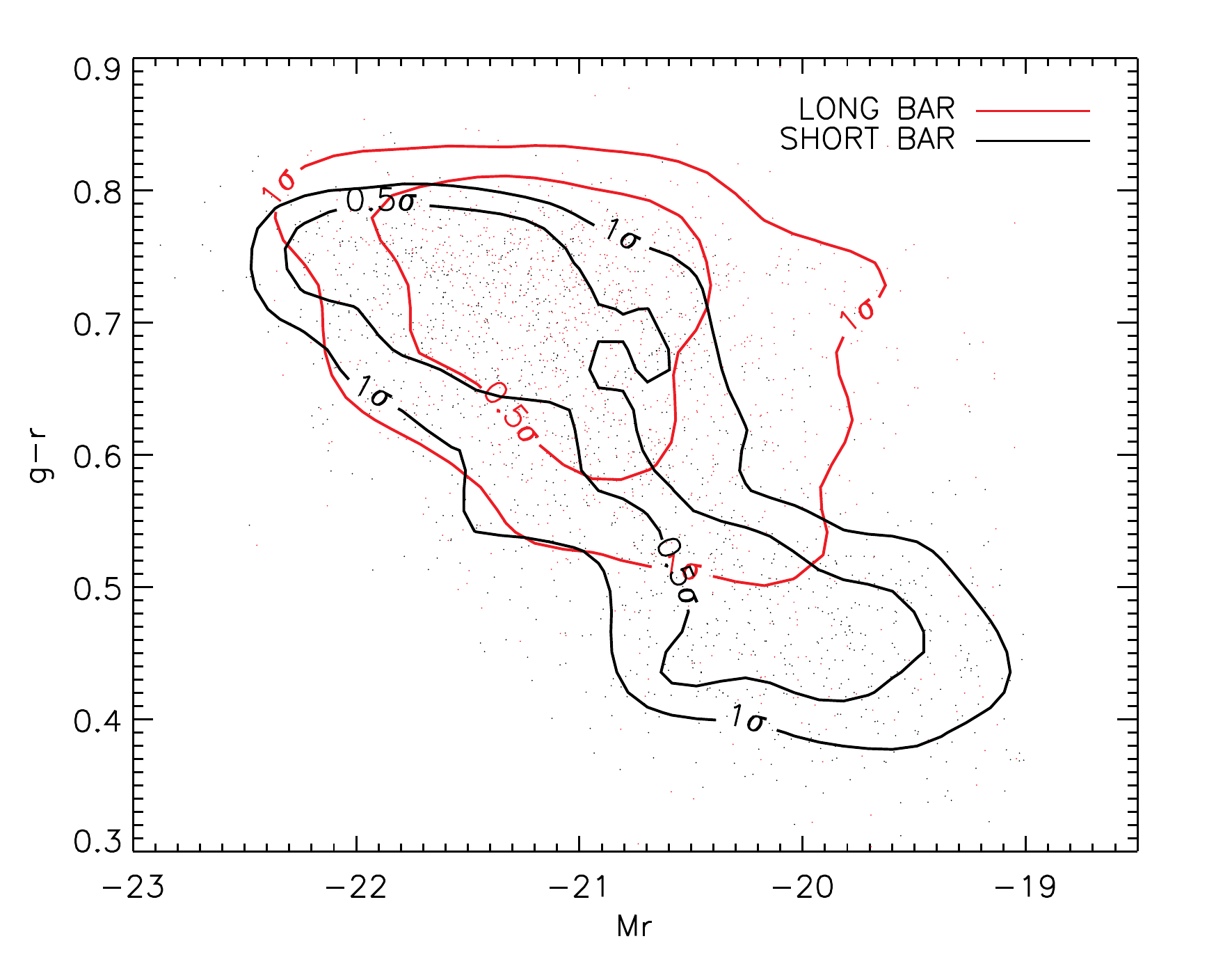}
\caption{Color-magnitude diagram for {\em barred} galaxies. The 0.5-$\sigma$ and 1-$\sigma$ contours are shown for comparison. Long-barred galaxies are brighter and redder.
}
\label{cmdbar}
\end{figure}

We divide barred galaxies into two subgroups using the relative bar length cut of $L_{\rm Bar} / 2 R_{\rm pet} = 0.5$. This cut yields roughly the same number of long-barred (1,191) and short-barred (1,231) galaxies. The two subgroups occupy different regions in the CMD (Figure~\ref{cmdbar}). Recalling the distribution of unbarred galaxies shown in Figure~\ref{cmd1}, there is a gradual transition from unbarred, through short-barred, to long-barred galaxies in this diagram. This distinction on the CMD according to bar length has recently been reported by Hoyle et al. (2011).

\begin{figure}
\centering
\includegraphics[width=0.45\textwidth]{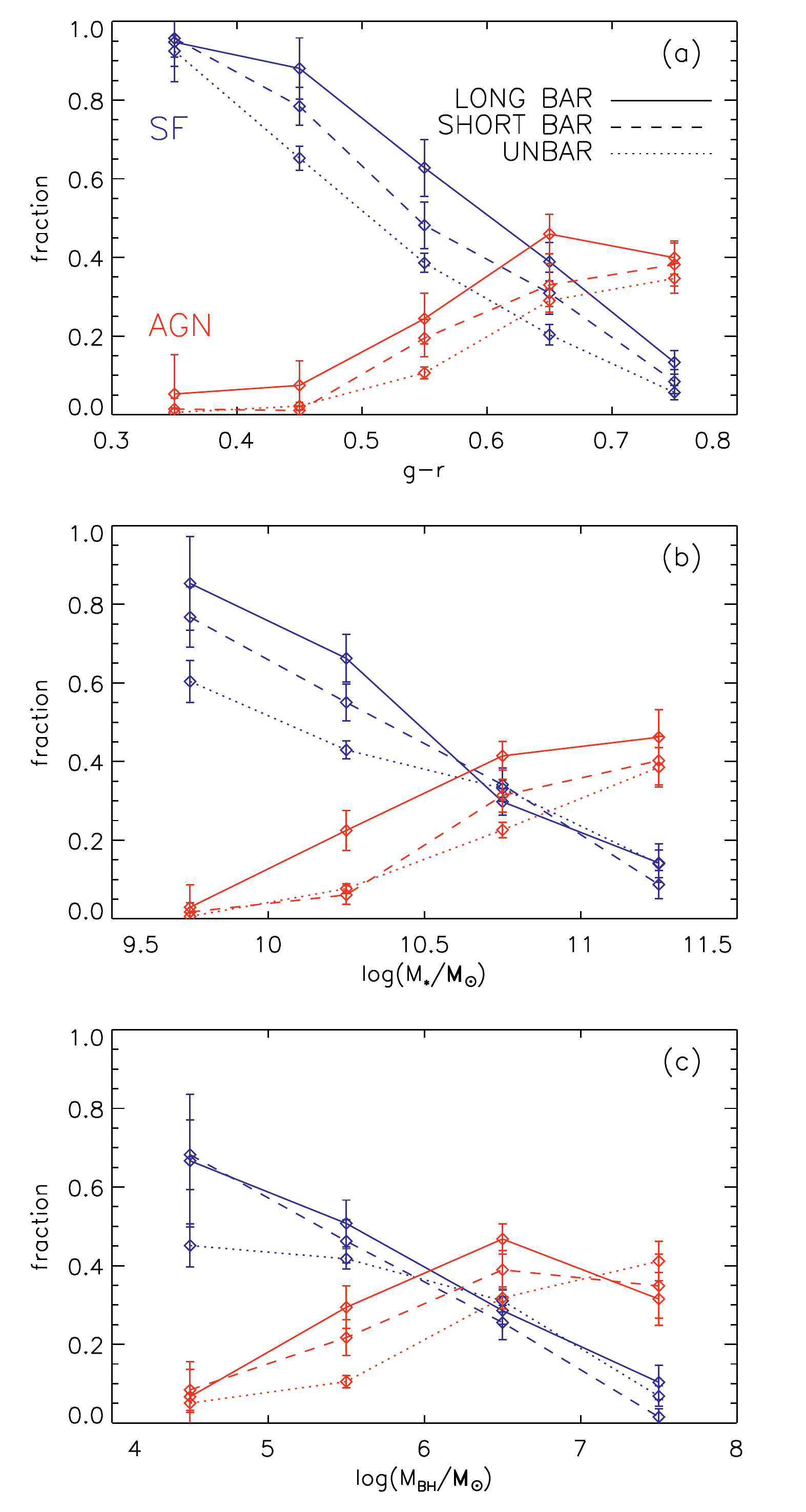}
\caption{The effect of bar length on the galaxy fractions with central star formation and AGN activity. The blue and red lines represent central star formation and AGN, respectively. Poisson errors are denoted as error bars.}
\label{frac2}
\end{figure}

According to the BPT diagnostics discussed in Section 3.2, 34.3\% of {\em long-barred} galaxies are classified as ``significantly'' star-forming galaxies. 36.9\% are AGN, and the rest (28.8\%) do not show all four emission lines above our cut. Short-barred galaxies are classified into star-forming galaxies (43.3\%), AGN (22.3\%), and the rest (34.4\%). We have already discussed in Section~\ref{baronnuc} that these simple-minded fractions are insufficient for describing the bar effects.

We plot the galaxy fractions with central star formation and AGN for barred galaxies as a function of galaxy properties in Figure~\ref{frac2}. The overall trends are the same as in Figure~\ref{frac1}. Long-barred galaxies show higher fractions of both central star formation and AGN, compared with short-barred galaxies. There is a gradual increase of both central star formation and AGN from unbarred galaxies, through short-barred galaxies, to long-barred galaxies.

We have also compared specific H$\alpha$, [OIII], and [NII] emission luminosities for long and short barred galaxies, as shown in Figures~\ref{SFflux1} and \ref{AGNflux1}. Short-barred galaxies are very similar to unbarred galaxies (not present in this figure) in these diagrams, and long-barred galaxies show more enhanced emission luminosities than barred galaxies. 

In conclusion, we see a monotonic effect of bar length on emission luminosities as well as on the galaxy fractions with central star formation and AGN activity.

\section{Discussion}
We have investigated the effect of bars on central star formation and AGN activity using a large sample of galaxies based on the SDSS DR7. With our selection criteria (Table \ref{tab:sample}), 6,658 late-type galaxies are selected for our sample, and among them 2,442 (36\%) galaxies are visually classified as barred galaxies. Our visual inspection may have missed galaxies with a weak bar, in which case our results would be more applicable to strong-bar galaxies.

We tried our best to avoid sampling biases. First, we adopted volume-limited sampling approach. However, our final sample after visual inspection exhibits a mild mass bias, mainly because visual inspection tends to reject smaller galaxies at higher redshifts. Hence, we restricted ourselves to the small redshift range adopted for this study to minimize such a bias. Second, we inspect bar effects for fixed stellar mass and/or black-hole mass in order to avoid any complex side-effect from such biases. To test the impact of the bias from the size effect, we have performed all tests in this paper by dividing our sample into two equal-number groups by redshift cut and found that our results are robust against the size bias. Our analysis is further subject to uncertainties in the AGN/SF classification scheme. Different classifications will result in different samples, and conclusions can be fairly dependent on the details in the classification scheme. For example, the current, incomplete understanding on transition objects adds uncertainties, and AGN-like optical emissions can mislead the analysis (Sarzi et al. 2010; Cid Fernandes et al. 2011). We found our results are pretty much the same with original ones when we applied stricter criterion ($W(H\alpha) >$ 3\,\AA) for AGN (e.g. Figure~\ref{grid22}; Figure~\ref{AGNflux1}; Figure~\ref{frac2}).


Barred galaxies are generally earlier in morphology, optically redder and more massive than unbarred galaxies. Consequently, the criteria used to construct an initial sample is a key issue for studying barred galaxies. Since galaxy colors and concentration indices are correlated with galaxy morphology, they are often used for initial sample selection. However, these parameter-based selections can miss a significant number of barred galaxies, for example, those with earlier morphology. 

A large-scale bar can be a channel of gas inflow, and the infalling gas activates both central star formation and AGN under certain conditions. Our results provide quantitative support for the theoretical predictions of the bar fueling scenario in central star formation and AGN. Central star formation is common in blue and/or low-mass galaxies and the presence of a bar does not seem to affect much the incidence of intensity of such an activity. On the other hand, bars can boost significantly the small fraction of red spiral galaxies showing signs of star formation, leading to central starbursts that are more intense than what observed in blue spirals.

Most of the barred galaxies with blue colors and/or low mass have a shorter bar. We have found that central star formation is more pronounced with increasing bar length. The simulations and observations expect longer bars are stronger, particularly in their ability to supply gas to the circumnuclear region. Therefore, the result that bars activate central star formation more in redder galaxies might have been caused by the fact that redder galaxies tend to have a longer bar.

The amount of gas could be another reason for the elevated bar effect on redder star-forming galaxies. Galaxies with bluer optical colors are expected to have sufficient gas, and that may be enough to maintain star formation. However, as galaxy color gets redder, the amount of gas is significantly reduced, and so central star formation is naturally declined on redder galaxies. Therefore, infalling gas through large-scale bars may have a larger effect on gas-deficient red galaxies.

There may be other reasons that bar effects on central star formation are not visible in blue galaxies, for instance due to fueling time-scale arguments on the bar. ``Fueling time-scale'' in this study means the time it takes for gas to travel to inner kiloparsec regions of a galaxy through bar effects. We would not expect that bars younger than their associated fueling time-scales have an impact on central star formation and AGN. However, it is not easy to estimate the age of a bar from observation. According to numerical calculations, bars grow longer and stronger with dynamical age by losing their pattern speed (e.g. Sellwood 1981; Athanassoula 2003). If fueling time is, for example, linearly proportional to bar length while bar length grows with time much more slowly, both of which are reasonable, then long-barred galaxies would have a better chance of having a larger bar age than the associated fueling time and experiencing central star formation. Likewise, blue galaxies with a short bar would not have enough time to supply gas into the nuclear region.

AGN are found mainly in red and massive galaxies, but bar effects on AGN are mainly found in galaxies with relatively blue colors and low black-hole mass. Contrary to central star formation, bar effects in AGN are not higher in galaxies with earlier morphology. We attempt to explain this in terms of the effect of central mass concentration on bar strength. Numerical simulations suggest that central mass concentration such as a massive black hole can significantly weaken bar structures (Friedli \& Pfenniger 1991; Hasan, Pfenniger, \& Norman 1993; Norman, Sellwood, \& Hasan 1996; Athanassoula, Lambert \& Dehnen 2005). Observationally, Das et al. (2003) also claimed that a central mass concentration could affect bar strength. However, it is not so simple. It may be reasonable to assume that a larger black hole has a negative effect on the gas infall (``negative black hole mass effect''). But we also found that early-type spirals generally have a longer bar (``positive longer bar effect''), as well as a larger black hole, which is expected to have a positive effect on the gas infall. In this regard, the bar effect in early-type spirals is complex and the net effect is a result of competition between these two. 

We focus on the results on {\em early-morphology spiral galaxies}. Relative bar length in galaxies with {\em central star formation} tightly correlates with black-hole mass (Figure~\ref{bleng1}-(c)). In this subgroup of galaxies, the positive longer bar effect seems to be winning. On the contrary the negative black hole mass effect seems to be winning in AGN host galaxies probably because bar length does not keep up with black-hole mass in AGN-host galaxies, as shown in Figure~\ref{bleng1}-(c). Again, the difference between star-forming and AGN spirals may indicate that they are in different stages of bar evolution.

A simpler possibility is that there is little or no gas supply through the bar in AGN-host galaxies which have earlier morphology. Sheth et al. (2005) found a significant fraction of barred galaxies having very little molecular gas within the bar and nuclear regions, among galaxies having earlier morphology. They suggested that the gas within the bar has already been consumed by star formation. If AGN having earlier morphology are in the post-starburst phase, gas might have already been consumed by central star formation, and there would be no more gas to feed AGN.

In conclusion, bar effects on central star formation and AGN activity are clearly visible and as expected. But they are also complex. A careful sampling strategy and degeneracy-breaking analysis are necessary to find them. This has become possible with a large galaxy database. Theory now needs to explain the physics of the empirical findings.

\section*{acknowledgments}
We are grateful to the anonymous referee for constructive comments that greatly improved the clarity of the paper. We would like to thank Martin Bureau and Min-Su Shin for taking the time to give valuable advice. We also acknowledge Marc Sarzi and Johan Knapen for stimulating discussion. SKY acknowledges support from the National Research Foundation of Korea to the Center for Galaxy Evolution Research, and from Doyak grant (No. 20090078756). This project made use of the SDSS optical data.

\end{document}